# Transport critical current densities and *n* factors in mono- and multifilamentary MgB$_2$/Fe tapes and wires using fine powders

H.L. Suo, P. Lezza, D. Uglietti, C. Beneduce, V. Abächerli and R. Flükiger

*Abstract*— Mono- and multifilamentary MgB$_2$/Fe tapes and wires with high transport critical current densities have been prepared using the powder-in-tube (PIT) process. The fabrication details are described. The effect of powder grain sizes and recrystallization temperature on $j_c$ has been investigated. At 25K and 1 T, $j_c$ values close to $10^5$ A/cm$^2$ were measured, while $j_c$ of $10^6$ A/cm$^2$ were extrapolated for 4.2K/0T in our monofilamentary tape. MgB$_2$/Fe tapes exhibit high exponential *n* factors for the resistive transition: $n \approx 80$ and 40 were found at 5 T and 7 T, respectively. The highest transport $j_c$ values obtained so far in MgB$_2$/Fe wires with 7 filaments were $1.1 \times 10^5$ A/cm$^2$ at 4.2 K and in a field of 2 T, which is still lower than for monofilamentary tapes. The function $Fp \propto b^p \cdot (1-b)^q$ has been established over the whole field range, and exhibits a maximum at $F_p \cong 0.18$. Improved deformation and recovering processing is expected to lead to higher $j_c$ values.

*Index Terms*— Mono- and multifilamentary MgB$_2$ superconducting tapes and wires; transport critical current densities; Fine MgB$_2$ powder; *n* factors; Pinning force

## I. INTRODUCTION

The recently discovered superconductor MgB$_2$ with $T_c = 39$ K [1] is characterised by weak link free grain boundaries [2] and low material cost, and is thus promising for potential application at 25 K. The feasibility of MgB$_2$ wires or tapes with high critical current density using the Powder-In-Tube (PIT) method has been demonstrated, either with [3]-[10] or without [11], [12] recrystallization after deformation. The $j_c$ values of these tapes at 4.2 K, 0 T are in the range of $\sim 10^5$-$10^6$ A/cm$^2$. We have previously reported [3], [4] on the fabrication of highly dense monofilamentary MgB$_2$ tapes with large $j_c$ values using Ni and Fe as sheath materials [3], [4], [6].

Manuscript received August 5, 2002. This work was supported by the Fond National Suisse de la Recherche Scientifique

Hongli Suo, P. Lezza, D. Uglietti, V. Abächerli and R. Flükiger are with Département de Physique de la Matière Condensée, Université de Genève, 24 Quai Ernest-Ansermet, Genève 4, CH-1211, Switzerland (telephone: +41-22-7026578, e-mail: Hongli.Suo @physics.unige.ch).

C. Beneduce is with Bruker Biospin, Magnetics Division, Industriestrasse 26, Ch-8117 Fällanden, Switzerland.

In this work, we restrict our study on Fe as a sheath material and investigate more deeply the effect of the grain size of the starting powder and the annealing temperature on the properties of MgB$_2$ tapes. We found a significant enhancement of the *n* factors in our monofilamentary tapes using finer starting powder. The results on MgB$_2$/Fe multifilamentary wires and tapes are also presented.

## II. EXPERIMENTAL

The fabrication of the MgB$_2$/Fe tapes was previously described [3], [4], [13]. In the present work, we analyzed the effect of various powder grain sizes obtained by ball milling for 2, 3, 14 and 100 hours, respectively. After drawing to 2 mm diameter, the wires were cold-rolled to tapes. Annealing was performed between 920°C and 980°C for 0.5h in Ar atmosphere.

## III. RESULTS

### A. $j_c$ Values in Monofilamentary MgB$_2$/Fe Tapes

*1) Thickness and $j_c$ in As-rolled and Annealed Tapes*

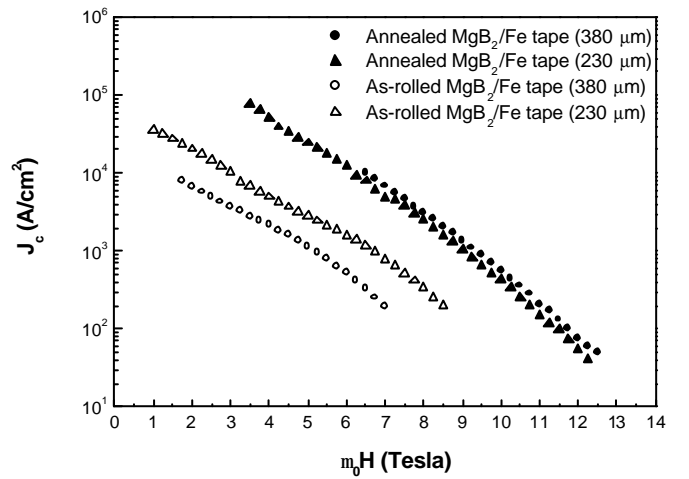

Fig. 1. Field dependence of the transport $j_c$ values at T = 4.2 K in both as-rolled and annealed MgB$_2$/Fe tapes with different thicknesses.

We fabricated the tapes with thicknesses of 380 and 230 μm, corresponding to a filling factor of 28% and 25%, respectively. The field dependence of $j_c$ in these MgB$_2$/Fe tapes, both before and after the annealing, is shown in Fig. 1. A comparison at 2 T shows that the thinner as-rolled $MgB_2$/Fe tape (230 μm) yields higher current densities, reaching $2 \times 10^4$ A/cm$^2$ at 4.2 K (unannealed). This result indicates that the tape with lower filling factor corresponds to a higher $j_c$ value, this confirming the observation of Grasso [11] and showing that the filling factor is an important parameter.

The final recrystallization anneal caused a densification, then leading to a strong enhancement of $j_c$ to $8 \times 10^4$ A/cm$^2$ in a field of 3.5 T for 230 μm thick Fe/$MgB_2$ tape and $10^4$ A/cm$^2$ in a field of 6.5 T for the 380 μm thick Fe/$MgB_2$ tape. After recrystallization, similar $j_c$ vales were obtained for both thicknesses, thus suggesting that the final density inside the filament is quite similar.

Quenching occurred in all our samples above a given current, which was higher in 230 μm thick $MgB_2$/Fe tape than in the 380 μm $MgB_2$/Fe one. This difference is thought to be due to the smaller core thickness of the thinner $MgB_2$/Fe tape (70 μm compared to 160 μm for the 380 μm thick $MgB_2$/Fe tape). Further reduction of the filament thickness is expected to improve the thermal stability of $MgB_2$/Fe tapes and wires.

*2) Effect of Grain Size of Starting Powder on $j_c$*

We have previously reported [13] that after reducing the $MgB_2$ grains to a micrometer size by ball milling, both the critical current density, $j_c$ and the irreversibility field, $\mu_0 H_{irr}$ were enhanced, while the upper critical field, $\mu_0 H_{c2}$, remained unchanged. In the present work we performed ball milling in an agate mortar containing agate balls for times t = 2-100 h. The various powder sizes in Fig. 2 are represented by the peak values in the size distribution as obtained by granulometry: (a) after 2 hours the two observed peaks at 3 and 30 μm, respectively, are characterised by 3/30 μm, and similarly, (b) 3 hours: 1.5/10 μm, (c) 14 hours: 1/7 μm and (d) 100 hours: 1 μm.

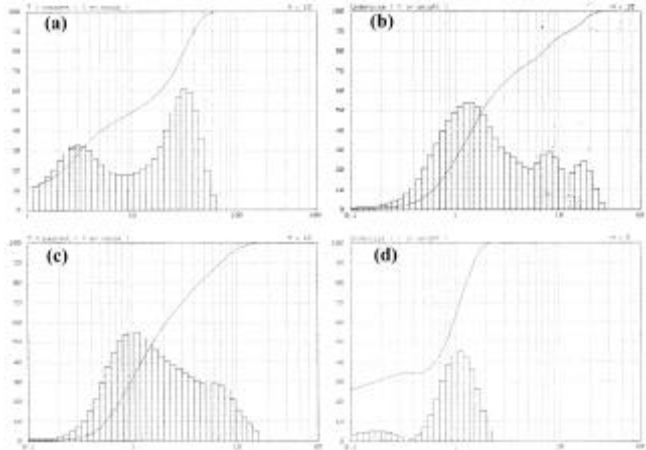

Fig. 2. Distributions of powder grain sizes: (a) 2 h: powder 3/30 μm; (b) 3 h: powder 1.5/10 μm; (c) 14 h : powder 1/7 μm; (d) 100 h: powder 1 μm

We prepared Fe/$MgB_2$ tapes using the powders (a) to (d) by the same deformation processing and heat treatment. Fig. 3 compares the $j_c$ values in these tapes annealed at 950°C. For comparison, the $j_c(H)$ curves of an annealed $MgB_2$/Fe tape produced directly using as-purchased powder with a wide size distribution centred at around 60 μm are also shown in the same figure. For the initially coarse-grained powder, the $j_c$ is markedly lower than that of ball milled tapes. In the tape with powder (a) 3/30 μm, we obtained the highest $j_c$ value of $10^4$ A/cm$^2$ at 4.2 K and in a field of 6.5 T. Extrapolating the field dependence of $j_c$ in this tape yielded self-field values close to 1 MA/cm$^2$. At T = 25 and 30 K, $j_c$ values well above $10^4$ A/cm$^2$ were obtained at fields of 2.25 and 1.0 T, respectively. With the reduction of the grain size of starting powder, the $j_c$ values slowly decreased, the lowest $j_c$ values being observed in the tape based on powder (d) 1 μm. In principle, it was expected that finer powders would lead to higher $j_c$ values due to the enhanced interface between grains. The actual results suggest a possible influence of impurities at the powder surface, as a consequence of the grinding procedure. The powder with the smaller grain size have considerably larger total grain surface, which makes the grains more active thus absorbing more oxygen during high temperature annealing, therefore resulting in a degradation of the transport $j_c$ values.

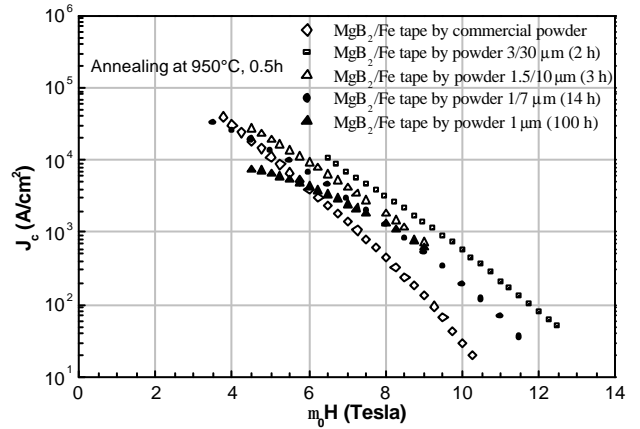

Fig. 3. Field dependence of the transport $j_c$ values at T = 4.2 K in annealed $MgB_2$/Fe tapes prepared by different powders.

*3) Effect of Annealing Temperature on Transport $j_c$*

We studied the effect of annealing temperatures between 920°C and 980°C on $j_c$ values in tapes prepared by powder (b) 1.5/10 μm. As shown in Fig. 4, a higher $j_c$ value is found when lowering the temperature from 980°C to 920°C. At 920°C, we obtained the same $j_c$ values in this tape as for the tape with by powder (a) 3/30 μm, annealed at 950°C. This result suggests that the optimum annealing temperature could differ for $MgB_2$ tapes based on different powder grain sizes: for tapes produced by smaller grain size, a lower annealing temperature has proven to give better results (Fig. 4). Our experiments show a good reproducibility of the high $j_c$ values of tapes produced in different ways.

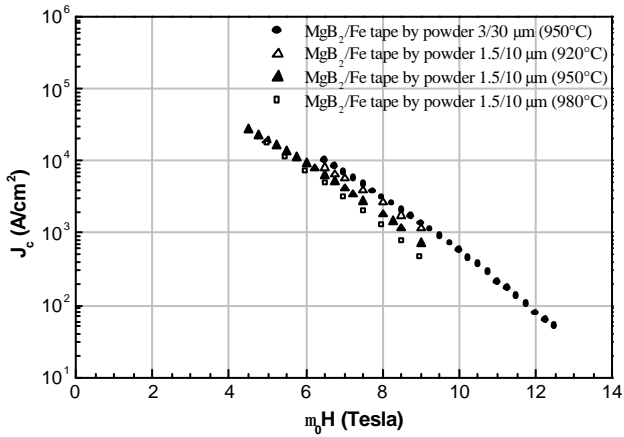

Fig. 4. Field dependence of the transport $j_c$ values at T = 4.2 K in MgB$_2$ tapes annealed at different temperatures.

*4) Exponential n factors in Monofilamentary MgB$_2$/Fe Tapes*

To analyze the possibility of MgB$_2$ conductor to work in the persistent mode, we studied the exponential $n$ factors of the tapes. The logarithmic $E$-$J$ curves measured at 4.2 K can be reasonably well approximated by a local power-law, with the electric field criterion $E_c = 10^{-6}$ V/cm: $(E/E_c) \approx (j/j_c)^n$.

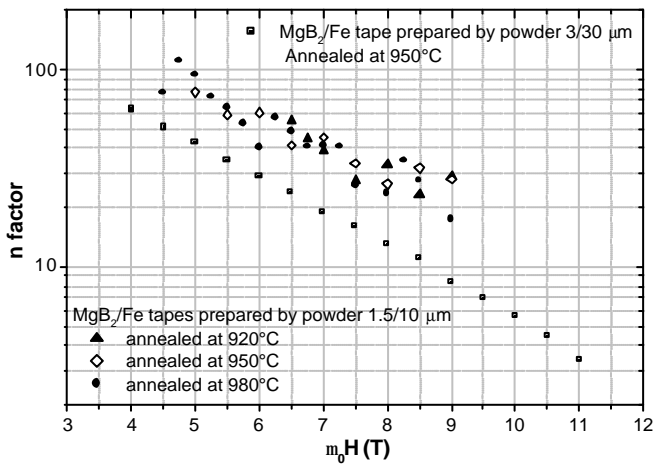

Fig. 5. Field dependence of the exponential n factors in annealed MgB$_2$/Fe tapes.

By fitting this relation to our data in the electric field range of 0.5 µV/cm < E < 5 µV/cm, we obtained $n$ factors shown in Fig. 5 for the tape prepared by powder (a) 3/30 µm. The factor is determined to be ~ 30 at 6 T, and decreases exponentially to 10 at 8.5 T. By fitting our data in the range of 0.1 µV/cm < E < 1 µV/cm, we obtained the factors for the tapes fabricated by powder (b) 1.5/10 µm, also shown in Fig. 5. A strong enhancement of $n$ factors was observed: the $n$ factors of the three tapes annealed at 980, 950, 920°C are similar, being ~ 80 at 5 T and also decreasing exponentially to ~ 40 at 7 T. This improvement of $n$ factors may be related to the smaller grain sizes used for those tapes and thus to the corresponding higher local homogeneity. The high $n$ factors open the possibility to use MgB$_2$ magnets in the persistent mode for fields up to 7 T at 4.2 K and to 2.5 T at 25 K.

### B. Multifilamentary Wires and Tapes

We have previously reported [14] the fabrication and superconducting properties of multifilamentary MgB$_2$/Fe wires. The highest $j_c$ values were obtained in MgB$_2$/Fe square wires (1.7 mm dimension) with 7 filaments produced by two-axial rolling, followed by annealing at 950°C. More recently, in order to decrease core thickness and improve thermal stability, we have developed a new configuration of multifilamentary tape with 9 filaments. The preparation consists of packaging thin MgB$_2$/Fe single tapes into rectangular Fe tube (Width: 9 mm, Thickness: 5 mm) and repeating the same deformation (two-axial rolling) procedure. Two recovery annealings at 600°C for 1 hour were performed for softening the Fe sheath which showed considerably work hardening during rolling. The annealed wire was deformed again by two-axial rolling to final tape thickness of 0.9 mm. The multifilamentary tape has very uniform transversal sections, as can be seen from the SEM micro-graphs of polished cross-sections of this tape (Fig. 6).

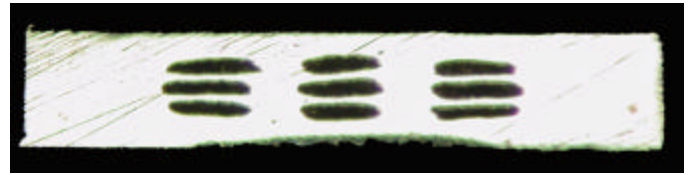

Fig. 6. Transversal cross-section of multifilamentary MgB$_2$/Fe tape with 9 filaments (thickness: 0.9 mm)

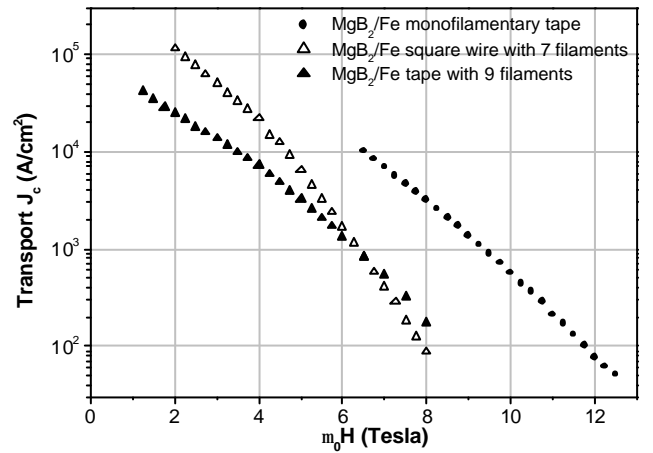

Fig. 7. Transport critical current densities at T = 4.2 K as function of applied field in annealed MgB$_2$/Fe multifilamentary wires. For comparison, the transport $j_c$ curve of a monofilamentary MgB$_2$/Fe tape is also shown

Fig. 7 summarizes the $j_c$ values at 4.2 K as function of applied field for our multifilamentary MgB$_2$/Fe wires (1.7 mm dimension) and tapes (0.9 mm thickness). The $j_c$ value in the MgB$_2$/Fe square wire with 7 filaments was $1.1\times10^5$ A/cm$^2$ at 4.2 K and 2 T. Quenching was also observed on the present multifilamentary wires, but the measurements could be extended to lower field values. The estimated self-field value of $j_c$ at 4.2 K in this square wire was close to $4 \times 10^5$ A/cm$^2$. The cross over of $j_c$ shown in Fig. 7 between the two multifilamentary configurations has not been fully understood and may be due to different states of densitiy and texturing.

Compared with the MgB$_2$/Fe monofilamentary tapes, the $j_c$ values in both multifilamentary conductors are substantial lower, which might be due to the poor connectivity of grains (confirmed by SEM), indicating a lower density. In addition, the effect of these intermediate anneals on the 9-filaments tape may have had a negative influence. The improvement of thermal stability, deformation processing as well as the optimization of intermediate anneals appear a challenge for the future development of multifilamentary MgB$_2$/Fe tapes.

### C. Pinning Force (in Monofilamentary Tapes)

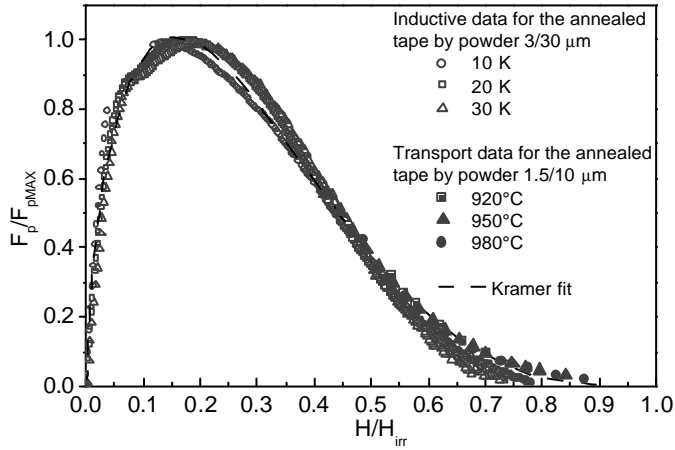

Fig. 8. Reduced pinning force at different temperatures for tape by powder 3/30 μm (open symbols). Close symbols refer to tapes by powder 1.5/10 μm. The dashed line is the Kramer law with exponents p=0.6 and q=3.2.

From $I_c(B)$ it is possible to calculate the pinning force, which can be fitted with the usual relation $F_p \propto b^p \cdot (1-b)^q$ where $b$ is the ratio between B and B*, B* being the irreversibility field. Unfortunately the range for the available transport measurements is restricted to high fields (4 T - 7 T), due to insufficient thermal stability. As a consequence it is quite difficult to fit the data to find $p$ and $q$ and to locate the position of the peak. Based on our previous results [13], [15] on the whole field range showing that the values of inductive and resistive $j_c$ are the same, we have performed an analysis using the inductive $j_c$ reported by Suo et al. [4] to establish an universal scaling law for MgB$_2$ monofilamentary tapes. The inductive data at 10, 20, 30K have been used to determine the exponents of the Kramer law. The irreversibility field has been estimated using the $j_c = 10^2$A/cm$^2$ criterion. The coefficients have been found to be $p$=0.6 and $q$=3.2 (quite different values respect to Nb$_3$Sn), and have been used to extrapolate the transport measurements to lower fields. Moreover B$_{irr}$ and F$_{pmax}$ have been calculated from the fit and the transport measurements have been plotted together with the inductive ones in the normalised plot shown in Fig. 8.

### IV. CONCLUSION

In conclusion, we reported on the preparation of mono- and multifilamentary MgB$_2$/Fe tapes and wires. We found that both the grain size of the starting powder and annealing temperature were determinant for getting higher $j_c$ values. The estimated self-field value of $j_c$ in annealed monofilamentary MgB$_2$/Fe tapes at 4.2 K was close to $10^6$ A/cm$^2$. An enhanced $n$ factor (e.g. ~ 80 at 5 T) was found for monofilamentary tapes prepared by finer starting powders, which is attributed to a higher local homogeneity. This result confirms the possibility to use MgB$_2$ tapes for persistent mode operation. Multifilamentary MgB$_2$/Fe wires and tapes have been prepared, showing lower $j_c$ values than the monofilamentary tapes. The estimated self-field values of $j_c$ at 4.2 K in these multifilamentary wires exceeded $4 \times 10^5$ A/cm$^2$. The function $Fp \approx b^p \cdot (1-b)^q$ has been established over the whole field range, and exhibits a maximum at $F_p \cong 0.18$. Improved deformation and recovering processing is expected to lead to higher $j_c$ values.